\documentclass[showpacs,preprintnumbers,amsmath,amssymb]{revtex4}

\usepackage{amsmath}
\usepackage{graphicx}
\usepackage{epsf}
\usepackage{psfrag}
\usepackage{epsfig}
\usepackage{graphics}

\begin{document}

\newcommand{\be}{\begin{equation}}
\newcommand{\ee}{\end{equation}}
\newcommand{\bq}{\begin{eqnarray}}
\newcommand{\eq}{\end{eqnarray}}
\newcommand{\dt}{\frac{d^3k}{(2 \pi)^3}}
\newcommand{\dtp}{\frac{d^3p}{(2 \pi)^3}}
\newcommand{\elemintreg}{\int_\Lambda \frac{d^4 k}{(2\pi)^4}}
\newcommand{\eps}{\epsilon}
\newcommand{\intx}{\int^{1}_{0} dx}
\newcommand{\pbruto}{\hbox{$p \!\!\!{\slash}$}}
\newcommand{\qbruto}{\hbox{$q \!\!\!{\slash}$}}
\newcommand{\lbruto}{\hbox{$l \!\!\!{\slash}$}}
\newcommand{\kbruto}{\hbox{$k \!\!\!{\slash}$}}

\title{{\bf Multiloop calculations with Implicit Regularization in massless theories}}

\date{\today}

\author{E. W. Dias$^{(a),(d)}$} \email []{edsondias@fisica.ufmg.br}
\author{A. P. Ba\^eta Scarpelli$^{(b)}$} \email[]{scarp1@des.cefetmg.br}
\author{L. C. T. Brito$^{(c)}$} \email[]{lcbrito@ufla.br}
\author{H. G. Fargnoli$^{(a)}$} \email []{hfargnoli@fisica.ufmg.br}

\affiliation{(a) Federal University of Minas Gerais - Physics
Department - ICEx \\ P.O. BOX 702, 30.161-970, Belo Horizonte MG -
Brazil}
\affiliation{(b) Centro Federal de Educa\c{c}\~ao Tecnol\'ogica - MG \\
Avenida Amazonas, 7675 - 30510-000 - Nova Gameleira - Belo Horizonte
-MG - Brazil}

\affiliation{(c) Universidade Federal de Lavras - Departamento de Ci\^encias Exatas \\
Caixa Postal 37, 37.200-000, Lavras, Minas Gerais, Brazil}

\affiliation{(d) Universidade Federal de S\~ao Jo\~ao del Rei - Campus Alto Paraopeba \\
CAP Rod. MG443, km 7 - 36420-000 - Ouro Branco/MG - Brazil}

\begin{abstract}

\noindent
We establish a  systematic way to calculate multiloop
amplitudes of infrared safe massless models with Implicit Regularization (IR),
with a direct cancelation of the fictitious mass introduced by the
procedure. The ultraviolet content of such amplitudes have a simple
structure and its separation permits the identification of all the
potential symmetry violating terms, the surface terms. Moreover, we
develop a technique for the calculation of an important kind of finite multiloop
integral which seems particularly convenient to use
Feynman parametrization. Finally, we discuss the Implicit Regularization of infrared
divergent amplitudes, showing with an example
how it can be dealt with an analogous procedure in the coordinate space.
\end{abstract}


\maketitle

\section{Introduction}

\indent Higher order calculations in massless theories can be
performed with the help of several techniques
\cite{Kotikov:1995cw}-\cite{Kastening:1999yh}. In this paper we show
that  IR is a good tool for treating typical multiloop massless
integrals. Among the reasons that justify the use of  IR, we
can cite two important ones. The first one is the fact that the
method works in the physical dimension of the theory, and this
avoids complications with theories which are sensible to dimensional
modifications \cite{'tHooft:1972fi}-\cite{Stockinger:2005gx}. The
second one is the simple algorithm that the  IR provides for
the identification of potentially symmetry violating terms: the
surface terms which come from basic divergent integrals with Lorentz
indices \cite{edsonepj}, \cite{IR5}, \cite{IR6},
\cite{IR10}-\cite{IR12}.

It is just the second reason referred above that could bring some difficulties when one intends to apply the method
to massless models.  This is because the IR standard expansion, used to separate the  divergent
from the finite part of an integral, in this case can only be performed
with the introduction a fictitious mass; it turns out that the two parts are infrared divergent
(here we are talking about infrared safe  integrals). Of course, the parts  must be added in order to cancel this
mass; this is accomplished by means of a scale relation which introduces an independent mass parameter.
However, the basic divergent integrals
are a simple form of expressing the ultraviolet divergent content of the amplitude, since it does not ask for
an  explicit regularization \cite{cleber}.

On the other hand, this enforces the necessity of a method for performing the calculation of some pieces of
the finite part obtained by such expansion, or the cancelation of the fictitious mass will not be evident.
In this paper, we establish a systematic way to calculate multiloop integrals in  massless models.
In addition, we develop a technique for dealing with an important kind of ultraviolet finite integral which emerge from the expansion of
the integrand. We use a simple algebraic identity to put the integrand in a
convenient way to use Feynman parametrization. As a consequence, we obtain a direct
cancelation of the fictitious mass when the finite and divergent parts are put together.

It is well known that the use of a fictitious mass is not consistent when a genuine
infrared divergence is to be treated. For the sake of completeness, we briefly comment on a procedure to implicitly regularize
such kind of amplitudes, as it will be presented in \cite{WIP}.

This paper is organized as follows: in section two we overview the basics of Implicit Regularization;
the systematic calculation of multiloop massless integrals is presented in section three; we give
an example of a five-loop amplitude in section four; some comments on the Implicit Regularization of
genuine infrared divergent amplitudes are presented in section five;
the concluding comments are in section six. Detailed calculations of a typical finite part are presented in Appendix A and
the procedure for obtaining the scale relations is presented in the appendix B.

\section{Overview of Implicit Regularization}

Now, some important aspects of IR are
needed; the detailed discussion can be
found in the references \cite{IR3}-\cite{IR15}.
The condition is that the regularization, which is maintained implicit,
does not modify the  dimension neither the
integrand. So, a simple cutoff is a good choice, since the basic divergences will not be calculated, and
IR has a simple recipe to enforce symmetries.

For treating massless theories, the following steps should be carried out:

\begin{itemize}
\item the space-time and internal group algebra are performed and the amplitude is written as a combination of basic integrals;
\item taking into account that the integral is infrared safe, a fictitious mass $\mu^2$ is introduced. This must be done
because the characteristic expansion of Implicit Regularization breaks the integral in parts that are infrared divergent;
\item the expansion of the integrand is carried out until the divergent piece does not have the external momentum in the
denominator. The following identity is recursively used:
\be
\frac {1}{(p_i-k)^2-\mu^2}=\frac{1}{(k^2-\mu^2)}
-\frac{p_i^2-2p_i \cdot k}{(k^2-\mu^2) \left[(p_i-k)^2-\mu^2\right]},
\label{ident}
\ee
with $p_i$ a linear combination of the external momenta;
\item the divergent part is written in terms of basic divergent integrals of the type
\be
I_{log}^{(i+1)\,\mu_1 \cdots \mu_{\gamma}}(\mu^2)= \int_{k}^\Lambda
\frac{k^{\mu_1}\cdots k^{\mu_{\gamma}}}{(k^2-\mu^2)^\beta}
\ln^i{\left(-\frac{k^2-\mu^2}{\lambda^2}\right)},
\ee
with $\gamma=2\beta-4$, $\int_k \equiv \int d^4k/(2\pi)^4$ and where the $\Lambda$ is to indicate the presence of
an implicit regularization;
\item the basic divergent integrals with Lorentz indices are expressed as functions of basic divergent integrals
without indices and surface terms;
\item the finite part is calculated;
\item scale relations are used in order to write the basic divergent integrals in terms of a non-null, arbitrary mass
scale $\lambda^2$. The scale relation will allow an interplay between the finite and the divergent parts;
\item the limit $\mu^2 \to 0$ is taken.
\end{itemize}

\section{The calculation of multiloop integrals}

A complete discussion on the ultraviolet content of infrared safe massless models within the context of Implicit Regularization
is carried out in ref. \cite{cleber}. It has been shown that the divergent part of an amplitude
for such a theory can be completely displayed in terms of $I_{log}^{(i)} (\lambda^2)$, according to the definition
\begin{equation}
I_{log}^{(i)} (\mu^2) = \int_k^\Lambda \frac{1}{(k^2-\mu^2)^2}\ln^{i-1}{\left(-\frac{k^2-\mu^2}{\lambda^2}\right)}.
\label{I_logn}
\end{equation}
It has been also shown that after judiciously applying the identity (\ref{ident}), the divergent part of any multiloop amplitude will
be contained in a set of integrals of the type
\be
J_{\mu_1 \cdots \mu_r}= \int_{k}^\Lambda
\frac{k_{\mu_1}\cdots k_{\mu_r}}{(k^2)^\alpha(k-p^{\prime}_1)^2}
\ln^{i-1}{\left(-\frac{k^2}{\lambda^2}\right)},
\label{J}
\ee
with $p^\prime_1$ some linear combination of the external momenta and $2 \alpha \leq r+3$.
In the expression above, it will be necessary the introduction of a fictitious mass $\mu^2$.
A fictitious mass may always be introduced as long as the integral is infrared safe. This is
necessary when using IR to treat massless infrared safe ultraviolet divergent integrals,
because  the expansion of the integrand, as we will see below, breaks it
in two infrared divergent pieces. When a genuine infrared divergence appears this procedure can be problematic.
For such cases a new procedure within IR defining basic infrared divergent integrals is necessary \cite{WIP}, and
we will comment on this approach in section V.

There are other finite contributions to the finite part besides the ones that come from the integral of eq. (\ref{J}).
For some of them there is no analytic solution, but they have no problems with the limit $\mu^2 \to 0$. This fact can be simply showed
by proving that (\ref{J}) is infrared finite, since the whole amplitude is infrared finite.
So, we are interested here in the explicit calculation only of this part of the
finite content, because it is this one that will guaranty us, when  the ultraviolet divergent
part is considered together, the infrared finiteness of the amplitude. This justifies and shows the consistency
of IR for treating infrared safe massless models. As discussed in the paper \cite{cleber},
the calculation of higher order renormalization group functions can be completely carried out by knowing the coefficients of
the $I^{(i)}_{log}(\lambda^2)$'s which display the ultraviolet content of all infrared safe massless amplitudes.
Nevertheless, we must be secure that no new problem will emerge when the separation of these objects is performed.
This is the reason why we dedicate ourselves to establish a procedure for solving the finite part of integrals of the type (\ref{J}).

We now perform a complete calculation of a typical $n+1$-loop integral, which has the form of (\ref{J}):
\be
\label{eq: example}
I_\alpha^{(n+1)}=\int_k^\Lambda \frac{k_\alpha}{k^2(p-k)^2}\ln^n{\left(-\frac{k^2}{\lambda^2}\right)}
=\lim_{\mu^2 \to 0}\left\{  \int_k^\Lambda \frac{k_\alpha}{(k^2-\mu^2)
[(p-k)^2-\mu^2]}\ln^n{\left(-\frac{k^2-\mu^2}{\lambda^2}\right)} \right\}.
\ee
Carrying out the expansion of the integrand:
\bq
&&I_\alpha^{(n+1)}=\int_k^\Lambda  \frac
 {k_\alpha}{(k^2-\mu^2)}\ln^n{\left(-\frac{k^2-\mu^2}{\lambda^2}\right)}
\left\{ \frac {1}{(k^2-\mu^2)}  -\frac{p^2-2p\cdot k}{(k^2-\mu^2)^2}
+\frac{(p^2-2p\cdot k)^2}{(k^2-\mu^2)^2[(p-k)^2-\mu^2]}\right\} \nonumber \\
&& = 2 p^\beta I^{(n+1)}_{log\,\,\alpha \beta}(\mu^2) + \tilde  I_\alpha^{(n+1)}.
\eq
The last term, the finite part, is given by
\be
\tilde  I_\alpha^{(n+1)}= \int_k \frac{k_\alpha(p^2-2p\cdot k)^2}{(k^2-\mu^2)^3[(p-k)^2-\mu^2]}
\ln^n{\left(-\frac{k^2-\mu^2}{\lambda^2}\right)}.
\ee
We first turn our attention to the tensorial divergent integral, which is given by
\bq
&&I_{log \,\, \mu \nu} ^{(j)} (\mu^2) = \int_k^\Lambda \frac{k_\mu k_\nu}{(k^2-\mu^2)^3}\ln^{j-1}{\left(-\frac{k^2-\mu^2}{\lambda^2}\right)}=
\frac 14 \left\{ g_{\mu \nu}\int_k^\Lambda \frac{1}{(k^2-\mu^2)^2}\ln^{j-1}{\left(-\frac{k^2-\mu^2}{\lambda^2}\right)} \right. \nonumber \\
&& \left.+2(j-1) \int_k^\Lambda \frac{k_\mu k_\nu}{(k^2-\mu^2)^3}\ln^{j-2}{\left(-\frac{k^2-\mu^2}{\lambda^2}\right)}
-\int_k^\Lambda \frac{\partial}{\partial k^\nu}\left[\frac{k_\mu }{(k^2-\mu^2)^2}\ln^{j-1}{\left(-\frac{k^2-\mu^2}{\lambda^2}\right)}\right]
\right\}.
\eq
The procedure is recursively repeated for $I_{log \,\, \mu \nu}^{(i-1)}$, $I_{log \,\, \mu \nu}^{(i-2)}$ etc, until we obtain
\be
I_{log \,\, \mu \nu} ^{(j)} (\mu^2)= \frac{g_{\mu \nu}}{4} \sum_{i=1}^{j} \frac{1}{2^{j-i}}\frac{(j-1)!}{(i-1)!}
I_{log}^{(i)}(\mu^2)+ \mbox{surface terms}.
\label{CR}
\ee

Recall that we still have to deal with the fictitious mass, which in the limit $\mu^2 \to 0$ will give
infrared divergent pieces both in the ultraviolet divergent and finite parts. This problem is simply dealt
with by the use of regularization independent scale relations (they can be easily obtained with the help of a cutoff), which read
\begin{equation}
I_{log}^{(j)}(\mu^2)=I_{log}^{(j)}(\lambda^2)-b \sum_{k=1}^{j} \frac{(j-1)!}{k!}\ln^k{\left(\frac{\mu^2}{\lambda^2}\right)},
\label{scale}
\end{equation}
with $b=i/(4\pi)^2$, for arbitrary non-vanishing $\lambda$.
For infrared safe models a systematic cancelation of all powers of $\ln{\left(\frac{\mu^2}{\lambda^2}\right)}$ between
the ultraviolet divergent and finite parts finally crowns $\lambda$ a renormalization group scale.

We use equations (\ref{CR}) and (\ref{scale}) to write
\bq
&&I^{(n+1)}_{log\,\,\alpha \beta}(\mu^2)= \frac{g_{\alpha \beta}}{4} \sum_{k=0}^{n} \frac{1}{2^{n-k}}\frac{n!}{k!}
\left\{I_{log}^{(k+1)}(\lambda^2)-b \sum_{i=1}^{k+1} \frac{k!}{i!}\ln^i{\left(\frac{\mu^2}{\lambda^2}\right)}\right\} \nonumber \\
&&=\frac{g_{\alpha \beta}}{4} \sum_{k=0}^{n} \frac{1}{2^{n-k}}\frac{n!}{k!}
I_{log}^{(k+1)}(\lambda^2) -b \frac{g_{\alpha \beta}}{4} \frac{n!}{2^n}
\sum_{k=0}^n 2^k \sum_{i=1}^{k+1} \frac{1}{i!}\ln^i{\left(\frac{\mu^2}{\lambda^2}\right)}
\eq
After some algebra and the reorganization of the summations, we obtain
\be
I^{(n+1)}_{log\,\,\alpha \beta}(\mu^2)= \frac{g_{\alpha \beta}}{2}n! \sum_{i=1}^{n+1} \left\{\frac{1}{2^{n-i+2}}\frac{1}{(i-1)!}
I_{log}^{(i)}(\lambda^2)
-b \left(1-2^{i-n-2}\right)
\frac{1}{i!}\ln^i{\left(\frac{\mu^2}{\lambda^2}\right)}\right\}.
\ee
We see in the equation above that the second part, which is ultraviolet finite, diverges in the limit $\mu^2 \to 0$.
This part must be canceled by some contribution coming from the ultraviolet finite integral. We now turn ourselves
to this integral. There is an interesting trick that allows us to use the traditional Feynman parametrization for
solving this integral. We use the identity
\be
\ln{a}=\lim_{\epsilon \to 0} \frac{1}{\epsilon} \left(a^\epsilon -1\right)
\ee
to write
\be
\tilde I_\alpha^{(n+1)}= \lim_{\epsilon \to 0}\frac{1}{\epsilon^n} \sum_{k=0}^n (-1)^{n-k}\frac{n!}{k!(n-k)!}I^k_\alpha,
\ee
with
\be
I^k_\alpha=\frac{1}{(-\lambda^2)^{k \epsilon}} \int_k \frac{k_\alpha(p^2-2p\cdot k)^2}{(k^2-\mu^2)^{3-k\epsilon}[(p-k)^2-\mu^2]}.
\ee
We leave the detailed calculation of this finite part for the appendix. We select the term of order $n$ in $\epsilon$, since
it is the only one which contributes. The result is
\be
I^{k,n}_ \alpha=b p_\alpha \sum^{n + 1}_{i = 0}\left\{\left[1 - 2^{i - n - 2}\right]
\frac{1}{i!}\ln^{i}\left(\frac{\mu^2}{\lambda^2}\right)
-  (-1)^{n- i + 1}\frac{1}{2^{n - i + 2}}
\frac{1}{i!}\ln^{i}\left(-\frac{p^2}{\lambda^2}\right)\right\}(k\eps)^n.
\ee
Remembering that $I^{(n + 1)}_{\alpha}$ is given by
\begin{equation}
I^{(n + 1)}_{\alpha} = 2p^{\mu}I^{(n + 1)}_{log\,\mu\alpha}(\mu^2) + \lim_{\eps \rightarrow 0}
\frac{1}{\eps^n}\sum^{n}_{k = 0}(-1)^{n - k}\frac{n!}{k!(n -k)!} I^{(k)}_{\alpha}
\end{equation}
and using
\begin{equation}
\sum^{n}_{k = 0} (-1)^{(n - k)}\frac{k^n}{k!(n - k)!} = 1,
\end{equation}
we see the perfect cancelation of the dependence on the fictitious mass $\mu$. The final result is given by
\begin{eqnarray}
&&I^{(n + 1)}_{\alpha}=n! p_{\alpha}\left\{\sum^{n+1}_{i = 1}\left(\frac{1}{(i-1)!}
\frac{1}{2^{n - i + 2}}I^{(i)}_{log}(\lambda^2)
- b(-1)^{n - i + 1}\frac{1}{2^{n - i + 2}}
\frac{1}{i!}\ln^{i}\left(-\frac{p^2}{\lambda^2}\right) \right) \right. \nonumber\\
&& \left. +b\left[1 - \left(1 + (-1)^{n + 1}\right)2^{-(n + 2)}\right]
\right\}.
\label{ialpha}
\end{eqnarray}
The procedure adopted above can be applied to any integral of the type of (\ref{J}). The evaluation of multiloop amplitudes
in massless theories becomes simple, since the divergences and the finite piece necessary for the cancelation of the fictitious mass
are always originated in such kind of integral.
In the next section we show an example in which the only necessary result is given in equation (\ref{ialpha}).

\section{A multiloop amplitude}

\begin{figure}
\includegraphics{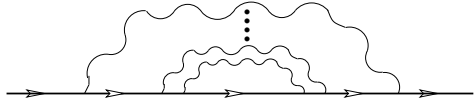}
\caption{$n$-loop nested selfenergy of spinorial QED.}
\label{self}
\end{figure}

We now perform, as an example, the calculation of a multiloop amplitude. Specifically, we will treat the n-loop nested selfenergy of spinorial
QED, as represented in the figure \ref{self}. With the subtraction of the subdivergences, we can write
\be
\Sigma^{(n)}(p)=\int_k^\Lambda \frac {\gamma^\rho \kbruto \tilde \Sigma^{(n-1)}(k)\kbruto \gamma_\rho}{k^4(p-k)^2},
\ee
where the tilde is to designate the finite part of the $(n-1)$th order graph. From the Lorentz structure, it is easy to see
that such an amplitude, whatever the order, may be displayed as $\Sigma^{(i)}(p)=\pbruto S^{(i)}(p^2)$, with $S^{(i)}(p^2)$ a scalar function
of $p^2$. So, we obtain
\be
\Sigma^{(n)}(p)=-2 \gamma^\alpha \int_k^\Lambda \frac {k_\alpha}{k^2(p-k)^2}\tilde S^{(n-1)}(k^2),
\ee
and again the tilde is used to designate the finite part. Now, let us suppose that
\be
\tilde S^{(n-1)}(k^2)= \sum_{i=0}^{n-1} a_i \ln^i\left(-\frac{k^2}{\lambda^2}\right).
\ee
So, we will get
\be
\Sigma^{(n)}(p)=-2 \gamma^\alpha \sum_{i=1}^{n} a_i I^{(i)}_\alpha,
\label{selfn}
\ee
according to the definition of the previous section. Consequently, from eq. (\ref{ialpha}), it is found
\be
\tilde S^{(n)}(k^2)= \sum_{i=0}^{n} b_i \ln^i\left(-\frac{k^2}{\lambda^2}\right).
\ee
From a simple calculation for the one-loop graph, we have
\be
\tilde S^{(1)}(k^2)= \frac{i}{(4 \pi)^2} \left[ \ln \left(-\frac{k^2}{\lambda^2}\right)-2 \right],
\ee
and by induction, it is proved that the result (\ref{selfn}) is correct. As a specific result, for
$n=5$ it is simple to obtain
\be
\Sigma^{(5)}(p)= -b^4\gamma^\alpha \left \{ \frac{1}{12} I^{(5)}_\alpha  - \frac{7}{6} I^{(4)}_\alpha
+ \frac{23}{4} I^{(3)}_\alpha - \frac{49}{4} I^{(2)}_\alpha + \frac{37}{4} I^{(1)}_\alpha \right \}.
\ee

\section{Comment on Implicit Regularization of Genuine Infrared Divergent Amplitudes}

The Implicit Regularization of infrared divergences is the subject of a work in progress \cite{WIP}. Here
we present the basics of the whole process. The essential steps which rendered Implicit Regularization
adequate in the case of ultraviolet divergences have their counterpart for infrared ones. Moreover, a
new scale appears, typically an infrared scale which is completely independent of the ultraviolet one.

First, let us consider the following ultraviolet divergent massless integral and its result within Implicit Regularization,
\be\label{ultradiv}
I = \int\frac{d^{4}k}{(2\pi)^{4}}\frac{1}{k^{2}(p-k)^{2}}
= I_{log}(\lambda^{2})-b\ln\left(-\frac{p^{2}}{\lambda^{2}}\right)+2b,
\ee
and then let us proceed with the following calculation:
\be\label{infra}
U = \int\frac{d^{4}k}{(2\pi)^{4}}\frac{1}{k^{4}(p-k)^{2}}.
\ee
By power counting $U$ is infrared divergent and ultraviolet finite. In order to be able to use all the
mathematics developed for ultraviolet divergent integrals we firstly note the following:
\be\label{kquatro}
\frac{1}{k^{4}}=-\int^{\Lambda} d^{4}u\ e^{iku}\int^{\Lambda}\frac{d^{4}z}{(2\pi)^{4}}\frac{1}{z^{2}(z-u)^{2}},
\ee
where $z$ and $u$ are configuration variables.

Note the strinking similarity between the above $z$ integral and equation (\ref{ultradiv}). We can thus write the result immediately:
\begin{eqnarray}
I(u^{2})&=&\int^{\Lambda}_{z}\frac{1}{z^{2}(z-u)^{2}}\nonumber\\
&=& \tilde{I}_{log}(\tilde{\lambda}^{-2})-b\ln\left(-u^{2}\tilde{\lambda}^{2}\right)+2b.\label{infrabdi}
\end{eqnarray}
But now $\tilde{I}_{log}(\tilde{\lambda}^{-2})$ is an infrared basic divergent integral and a scale relation
has been used in order to get rid of a fictitious length in favor of the infrared scale $l^2=1/ \tilde{\lambda}^2$.

Using this result and (\ref{kquatro}) in (\ref{infra}) we have
\begin{eqnarray}
&&U = -\int^{\Lambda}_{k}\frac{1}{(p-k)^{2}}\int^{\Lambda} d^{4}u\ e^{iku}I(u^{2})\nonumber\\
&&=-\frac{i}{(4\pi)^{2}}\int^{\Lambda}_{k}\int^{\Lambda} d^{4}u\int^{\Lambda} d^{4}x\frac{e^{i(p-k)x}}{x^{2}}e^{iku}I(u^{2})\nonumber\\
&&=\frac{i}{(4\pi)^{2}}\int^{\Lambda} d^{4}u\ \frac{e^{-ipu}}{u^{2}}\left(\tilde{I}_{log}(\tilde{\lambda}^{-2})\right.\nonumber\\
&&\left.-b\ln\left(-u^{2}\tilde{\lambda}^{2}\right)+2b\right)\nonumber\\
&&=\frac{1}{p^{2}}\left(\tilde{I}_{log}(\tilde{\lambda}^{-2})+b\ln\left(-\frac{p^{2}}{\bar{\tilde{\lambda}}^{2}}\right)+2b\right)\label{diviv},
\end{eqnarray}
with $\bar{\tilde{\lambda}}^2 \equiv \frac{4}{e^{2\gamma}}\tilde{\lambda}^{2}$, where $\gamma = 0,5772...$ is the Euler-Mascheroni constant.
This framework is similar to the Differential Renormalization of infrared divergences \cite{diff}.

\section{Conclusion}

This paper has been devoted to systematize the Implicit Regularization of massless models for any loop order.
For the infrared safe theories, a systematic way to evaluate multiloop Feynman integrals
in the context of Implicit Regularization has been presented. The ultraviolet content of
probability amplitudes have a simple structure and we can easily identify all the potential symmetry violating terms,
the surface terms. We have obtained a general
expression for the scale relations, which establish the interplay between the divergent and
finite parts of the amplitude, playing a fundamental role
in the elimination of terms which are dependent of the fictitious mass.
In addition, we have developed a technique for evaluating an important kind of finite Feynman integral which are typical of higher order
calculations with Implicit Regularization in massless
theories. We have extend the usual Feynman parametrization for
integrals which are not written in terms of rational functions of the momenta.
Through a simple example, we have exhibited the main elements of the procedure, showing how the finite part
and the scale relations work together in order to restore the infrared safety of the amplitudes.
Finally, we have discussed the Implicit Regularization of infrared divergent amplitudes, showing with an example
how it can be dealt with an analogous procedure in the coordinate space.

\section{Appendix A - The calculation of the finite part}

We carry out the detailed evaluation of $I^k$, which can be written, after Feynman parametrization and integration in $k$, as
\begin{equation}
I^{k}_{\alpha} = \frac{1}{(-\lambda^2)^{k\epsilon}}(A + B + C),
\end{equation}
where
\be
A=\frac{b}{(2-k\epsilon)}p_\alpha (-\mu^2)^{k\epsilon-2} \int_0^1 dx\,[p^2(1-2x)]^2x(1-x)^{2-k\epsilon}
\left(\frac{H^2}{(-\mu^2)}\right)^{k\epsilon-2},
\ee
\be
B=\frac{2bp^2}{(1-k\epsilon)(2-k\epsilon)}p_\alpha (-\mu^2)^{k\epsilon-1}
\int_0^1 dx\,(1-x)^{2-k\epsilon}x \left(\frac{H^2}{(-\mu^2)}\right)^{k\epsilon-1}
\ee
and
\be
C=-\frac{2b}{(1-k\epsilon)(2-k\epsilon)}p_\alpha (-\mu^2)^{k\epsilon-1}
\int_0^1 dx\, (1-x)^{2-k\epsilon}p^2(1-2x) \left(\frac{H^2}{(-\mu^2)}\right)^{k\epsilon-1}.
\ee
In the equations above, we have $b=i/(4\pi)^2$ and
\be
H^2=p^2x(1-x)-\mu^2.
\ee
Before considering the limit $\mu^2 \to 0$, some important manipulations have to be done. First, we observe that
\be
\frac{d}{dx}H^2=p^2(1-2x).
\ee
So, some integrations by parts are performed until the exponent of $H^2$ is $k\epsilon$. After this we can
write $H^2 \to p^2x(1-x)$ without problem. We will have, then,
\begin{eqnarray}
I^{k}_{\alpha} &=& -b\left[\frac{1}{1 - k\eps}\right]p_{\alpha}\left\{-\frac{1}{k\eps(2 - k\eps)}
\left(\frac{\mu^2}{\lambda^2}\right)^{k\eps} + \intx \left[\frac{(1 - x)}{k\eps} + (1 - 2x)\right]\left(-\frac{p^2 x}{\lambda^2}\right)^{k\eps}\right\}.
\end{eqnarray}
In this point, it is convenient to evaluate each term separately, and so, we label the first term by $\xi$ and the second one by $\zeta$
($I^k_\alpha=\xi+\zeta$). So, we can write
\begin{eqnarray}
\xi &=& b\frac{1}{2k\eps}\frac{1}{1 - k\eps}\frac{1}{1 - \frac{k\eps}{2}}\left(\frac{\mu^2}{\lambda^2}\right)^{k\eps}p_{\alpha}.
\end{eqnarray}
For small $\eps$, we can perform a binomial expansion in each term and after some algebra we obtain
\begin{eqnarray}
\frac{1}{2k\eps}\frac{1}{1 - k\eps}\frac{1}{1 - \frac{k\eps}{2}}
&=& \frac{1}{2k\eps}\sum^{\infty}_{j=0}\sum^{\infty}_{i = 0}(k\eps)^i\left(\frac{k\eps}{2}\right)^j\nonumber\\
&=& \frac{1}{2}\sum^{\infty}_{i = 0}\left[2 - \left(\frac{1}{2}\right)^i\right](k\eps)^{i - 1}.
\end{eqnarray}
Now we expand the term $\left(\frac{\mu^2}{\lambda^2}\right)^{(k\eps)}$, for $\epsilon \rightarrow 0$. After this, we have
\begin{equation}
 \xi = bp_{\alpha}\frac{1}{2}\sum^{\infty}_{i = 0}\left[2 - \left(\frac{1}{2}\right)^i\right](k\eps)^{i - 1}
 \left[1 + k\eps\ln\left(\frac{\mu^2}{\lambda^2}\right) + \frac{1}{2}(k\eps)^2\ln^2\left(\frac{\mu^2}{\lambda^2}\right) + \ldots\right].
\end{equation}
The coefficient of $(k\eps)^n$ will be given by
\begin{eqnarray}
& & \frac{1}{2}\left\{\left[2 - \left(\frac{1}{2}\right)^{n + 1}\right] + \left[2 - \left(\frac{1}{2}\right)^n\right]
\ln\left(\frac{\mu^2}{\lambda^2}\right)\right.\nonumber\\
&+& \left[2 - \left(\frac{1}{2}\right)^{n - 1}\right]\frac{1}{2}\ln^2\left(\frac{\mu^2}{\lambda^2}\right)
+ \ldots + \left[2 - \frac{1}{2}\right]\frac{1}{n!}\ln^n\left(\frac{\mu^2}{\lambda^2}\right)\nonumber\\
&+& \left. 2\frac{1}{(n + 1)!}\ln^{n + 1}\left(\frac{\mu^2}{\lambda^2}\right)\right\}.
\end{eqnarray}
Finally, the term of $\xi$ of order $n$ in $\epsilon$ can be written as
\begin{eqnarray}
\xi^{(n)} &=& b p_{\alpha}\sum^{n + 1}_{i = 0}\left[1 - (2)^{i - n - 2}\right]\frac{1}{i!}\ln^{i}\left(\frac{\mu^2}{\lambda^2}\right)(k\eps)^n.
\end{eqnarray}
For $\zeta$, we have
\begin{eqnarray}
\zeta = -b p_{\alpha}\intx dx \frac{1}{1 - k\eps}\left[\frac{1 - x}{k\eps} + (1 - 2x)\right]\left(-\frac{p^2 x}{\lambda^2}\right)^{k\eps},
\end{eqnarray}
which, in the $n^{th}$ order can be expressed as
\begin{eqnarray}
\zeta^{(n)} = -b p_{\alpha}\frac{1}{2}\sum^{n + 1}_{i = 0}(-1)^{n- i + 1}\frac{1}{2^{n - i + 1}}
\frac{1}{i!}\ln^{i}\left(-\frac{p^2}{\lambda^2}\right)(k\eps)^n.
\end{eqnarray}

\section{Appendix B - The scale relations}
\label{ap: A}
In this appendix we discuss the main steps which are necessary in order to obtain the scale relations at
$n^{th}$ order. As we have seen before, although the
integral (\ref{eq: example}) by itself is infrared finite, when the separation by means of the relation (\ref{ident}) is performed,
we are left with two infrared divergent parts. The scale relations are important
because they establish the connection between the finite and divergent
parts in order to make the limit $\mu^2 \to 0$ possible.

A typical basic logarithmic divergence at $(n+1)^{th}$ order in massless theories can be written, using a cutoff regulator, as
\begin{equation}
I^{(n + 1)}_{log}(\mu^2) = \int_k^\Lambda \frac{1}{(k^2 - \mu^2)^2}\ln^{n}\left(-\frac{k^2 - \mu^2}{\lambda^2}\right)
\end{equation}
and, going to the Euclidean space, it can be rewritten as
\begin{eqnarray}
I^{(n + 1)}_{log}(\mu^2) &=& b \int d(k^2)\frac{k^2}{k^2 + \mu^2}\ln^{n}\left\{\frac{k^2 + \mu^2}{\lambda^2}\right\}
= b\int^{\infty}_{\mu^2} dx \frac{x - \mu^2}{x^2}\ln^{n}\left(\frac{x}{\lambda^2}\right)\nonumber\\
&=& b \int^{\Lambda^2}_{\mu^2}\frac{dx}{x}ln^{n}\left(\frac{x}{\lambda^2}\right)
- \mu^2 b \int^{\Lambda^2}_{\mu^2}\frac{1}{x^2}\ln^n\left(\frac{x}{\lambda^2}\right)dx \quad \mbox{.}
\end{eqnarray}
The logarithmic scale relations can be obtained from the following difference
\begin{eqnarray}
I^{(n + 1)}_{log}(\mu^2) - I^{(n + 1)}_{log}(\lambda^2) &=& b\int^{\lambda^2}_{\mu^2}\frac{dx}{x}\ln^{n}\left(\frac{x}{\lambda^2}\right) -
b n! \sum^{n}_{i = 0} \frac{1}{i!}\ln^{i}\left(\frac{\mu^2}{\lambda^2}\right)
+ b n! \sum^{n}_{i = 0} \frac{1}{i!}\ln^{i}\left(\frac{\lambda^2}{\lambda^2}\right)\nonumber\\
&=& b\left\{-\frac{1}{n+ 1}\ln^{n + 1}\left(\frac{\mu^2}{\lambda^2}\right) -  n!\sum^{n}_{i = 0}
\frac{1}{i!}\ln^{i}\left(\frac{\mu^2}{\lambda^2}\right)\right\}\nonumber\\
&=& -b\sum^{n + 1}_{i = 0} \frac{n!}{i!}\ln^{i}\left(\frac{\mu^2}{\lambda^2}\right)
\end{eqnarray}
Finally, we write
\begin{equation}
I^{(n + 1)}_{log}(\mu^2) = I^{(n + 1)}_{log}(\lambda^2) - b\sum^{n + 1}_{i = 0} \frac{n!}{i!}\ln^{i}\left(\frac{\mu^2}{\lambda^2}\right).
\end{equation}
At one loop (n = 0), for example, we have:
\begin{equation}
I_{log}(\mu^2) = I_{log}(\lambda^2) - b\ln\left(\frac{\mu^2}{\lambda^2}\right).
\end{equation}
In two loops (n = 1), we obtain:
\begin{equation}
I^{(2)}_{log}(\mu^2) = I^{(2)}_{log}(\lambda^2) - b\left\{\frac{1}{2}\ln^2\left(\frac{\mu^2}{\lambda^2}\right) +
                                             \ln\left(\frac{\mu^2}{\lambda^2}\right)\right\} \quad \mbox{.}
\end{equation}

\vspace{2mm}
\noindent
{\bf Acknowledgements}

\vspace{2mm}
\noindent
A. P. Ba\^eta Scarpelli is grateful to CNPq for the financial support. The authors thank Marcos Sampaio, M. C. Nemes and
B. Hiller for relevant discussions on this work.

\end{document}